\documentclass[useAMS,usenatbib]{mn2e}
\usepackage{psfig, epsf, epsfig}
\title[Kinematics of globular cluster systems]{Kinematics of globular cluster systems 
and the formation of early-type galaxies}

\author[K. Bekki,  Michael A. Beasley, Jean P. Brodie,  Duncan  A. Forbes]
       {K. Bekki,${}^1$ Michael A. Beasley${}^2$,
        Jean P. Brodie${}^2$, and  Duncan A. Forbes${}^3$ \\
        ${}^1$School of Physics, University of New South Wales, Sydney, NSW 2052, Australia \\
        ${}^2$UCO/Lick Observatory, University of California, Santa Cruz, CA 95064 \\
        ${}^3$Centre for Astrophysics and Supercomputing, 
              Swinburne University of Technology,  
              Hawthorn, VIC 3122, Australia \\}
\date{Accepted 
      Received
      in original form 2001}

\pagerange{\pageref{firstpage}--\pageref{lastpage}}

\pubyear{2005}

\begin{document}

\maketitle

\label{firstpage}

\begin{abstract}
We numerically investigate the kinematic
properties of globular cluster systems (GCSs) in E/S0
galaxies formed from dissipationless merging of spiral
galaxies. The metal-poor globular clusters (MPCs) and
metal-rich clusters (MRCs) in the merger progenitors are
initially assumed to have spatial distributions consistent with
the Milky Way GC system.
Our principal results, which can be tested against observations,
are as follows. Both MPCs and MRCs in
elliptical galaxies formed from major mergers can exhibit
significant rotation at large radii ($\sim$20 kpc) due to
the conversion of initial orbital angular momentum into
intrinsic angular momentum of the remnant. MPCs show higher
central velocity dispersions than MRCs for most major
merger models. $V_{\rm m}/{\sigma}_{0}$ (where $V_{\rm m}$
and ${\sigma}_{0}$, are the GCS maximum rotational velocity and
central velocity dispersion of respectively)
ranges from 0.2--
1.0 and 0.1--0.9 for the MPCs and  MRCs respectively, within
$6R_{\rm e}$ for the remnant elliptical. For most merger
remnant ellipticals, $V_{\rm m}/{\sigma}_{0}$ of GCSs
within $6R_{\rm e}$ is greater than that of the field stars
within $2R_{\rm e}$. The radial profiles of rotational
velocities and velocity dispersions of the GCSs
depend upon the orbital configuration
of the merger progenitors, their mass-ratios, and the
viewing angle. For example, more flattened early-type
galaxies, formed through mergers with small mass ratios
($\sim$ 0.1), show little rotation in the outer MRCs. 
Two-dimensional (2D) velocity dispersion distributions of the
GCSs of merger remnant ellipticals are generally flattened
for both MPCs and MRCs, reflecting the fact that the GCSs have
anisotropic velocity dispersions. The 2D distributions of
line-of-sight-velocity of the GCSs in some remnant
ellipticals show minor-axis rotation, particularly for
MRCs. The kinematic properties of MPCs in merger remnant
ellipticals strongly resemble those of the surrounding dark
matter. This implies that the kinematics of MPCs in such
galaxies can be used to probe the kinematic properties of
their dark matter halos.
We discuss these results in the context of GC and galaxy
formation. We note a possible difference in the GC
kinematics between field and cluster Es and explain how GC
kinematics may help us understand the origin of S0
galaxies.
\end{abstract}


\section{Introduction}

The properties of globular cluster systems (GCSs)
in galaxies have long been considered as fossil records of galaxy
formation and subsequent evolutionary processes.
These properties have been discussed in various
different contexts of galaxy formation 
(Searle \& Zinn 1978; Forbes et al. 1997; Ashman \& Zepf 1998; Beasley et al. 2002;
Brodie et al. 1998;  C\^ote et al. 2000).
The specific frequencies ($S_{\rm N}$), colour bimodality, and 
structural properties of GCSs have received particular attention, 
generally focusing on origin of early-type (E/S0) galaxies.

The kinematics of GCSs, however, has not been extensively
studied. This is largely due to the practical difficulties of
obtaining large kinematic samples of globular clusters (GCs). 
However, GC kinematics
do offer important  insights into galaxy halo properties and 
galaxy formation processes. 
GCs in a galaxy are observable test particles
which trace the underlying gravitational potential beyond several effective
radii ($R_{\rm e}$), where the dark matter halo is thought to 
dominate the mass density. 
Therefore, the kinematics of GCs allow an estimation of 
the global mass distribution of a galaxy, including the dark matter halo.
In addition, kinematical differences between GCs with different chemical properties
can provide constraints on the formation processes of the
galaxy. This is analogous to the studies of stellar metallicities
and kinematics (i.e., orbital eccentricities) in the Galactic
halo used to assess the time scale of gravitational collapse of the Galaxy 
(e.g., Eggen, Lynden-Bell \& Sandage 1962).

Until recently, the lack of highly-multiplexing spectrographs
on large telescopes has meant that GC kinematics in only a handful of early-type galaxies 
have been studied in detail. 
They include M87 with 278 GC velocities (C\^ote et al.  et al. 2001; Cohen \& Ryzhov 1997; 
Mould et al. 1990), NGC 4472 with 263 (C\^ote et al. 2003; Zepf et al. 2000; 
Sharples et al. 1998; Mould et al. 1990), NGC 5128 with 215 GCs (Peng et al. 
2004a; Sharples 1988; Hesser, Harris \& Harris 1986) and NGC 1399 with 468 GCs 
(Richtler et al. 2004; Minniti et al. 1998; Grillmair et al. 1994).
Because of this small galaxy sample, it is difficult to draw any 
general conclusions.
For example, in M87 (C\^ote et al. 2001) the red GCs rotate about the galaxy minor axis, and the velocity 
dispersion is roughly constant with radius ($< \sigma >$ $\sim$ 397 km/s).
The outer blue GCs also rotate about the minor axis, but the inner ones rotate 
about the major axis. The velocity dispersion increases slightly with 
radius ($< \sigma >$ $\sim$ 364 km/s). 
In NGC 4472 (C\^ote et al. 2003), the red GCs do not rotate and the velocity dispersion is 
constant with radius ($< \sigma >$ $\sim$ 265 km/s). The blue GCs rotate 
about the minor axis, and also have a near constant velocity 
dispersion ($< \sigma >$ $\sim$ 342 km/s).

Various numerical simulations on the formation of GCs in galaxies 
have been used to better understand the observed properties of GCs (e.g., Weil \& Pudritz 2001; 
Bekki et al. 2002; Bekki \& Chiba  2002; Kravtsov \& Gnedin 2003;
Li et al. 2004). However, much of this work has focused on 
the origin of specific frequency ($S_{\rm N}$) of GCs in ellipticals, 
GC metallicity distributions, and color bimodality, 
rather than on GC kinematical properties.
Thus, at present, there are few model predictions that can be compared
with the above  observations.
The latest observations of extragalactic GCSs,  
based on multi-object spectrographs on large telescopes 
are providing unprecedentedly rich data on kinematical properties of the GCSs
(e.g., Richtler et al. 2004). 

Comparison between these kinematical data and 
numerical simulations not only allows investigation of 
the structure of dark matter halos (e.g., Peng et al. 2004a),
but will also help us to understand the formation processes of galaxies themselves. 

The purpose of this paper is to provide some observable predictions
of the kinematics of GCSs in early-type galaxies based upon dissipationless 
numerical simulations of galaxy mergers.
We adopt a reasonable set of parameters for the initial structural and kinematical
properties of GCSs in the merger progenitor spirals, and thereby
investigate dynamical evolution of the GCSs during galaxy merging.
The initial GCSs in merger progenitor spirals are assumed to be composed of 
old metal-poor GCs (MPCs) and old metal-rich clusters (MRCs), that 
are associated with their halos and bulges/thick disks respectively.
We investigate the structural and kinematical properties of
these MPCs and MRCs, but do not address the properties of young, metal-rich GCS 
that may be formed during galaxy mergers with gas (Bekki et al. 2002). 
These objects are expected to have quite different characteristics due to gas
dissipation (Bekki et al. 2002; Li et al. 2004), and appear to make
up only a small fraction of the metal-rich peak of GCSs (e.g., Brodie et al. 2004).
We investigate radial profiles of rotational velocities 
and velocity dispersions of GCSs in merger remnants (i.e., E/S0s)
and their dependencies on the physical parameters of
galaxy merging, such as the orbital configuration and the mass ratio of the merger
progenitors.

The plan of the paper is as follows: In the next section,
we describe our  numerical models  for
dynamical evolution of GCs in dissipationless galaxy mergers.
In \S 3, we present the numerical results
on structural and kinematical properties of MPCs and MRCs in
the merger remnants.
In \S 4, we discuss the derived numerical results in several
different contexts of galaxy formation and evolution,
such as cluster and field E formation. 
We summarise our  conclusions in \S 5.


\begin{table*}
\centering
\begin{minipage}{185mm}
\caption{Model parameters}
\begin{tabular}{cccccccccccccl}
{Model no.  
\footnote{PM and MM describe pair merger and multiple mergers, respectively.}} &
$m_{2}$   & 
$e_{\rm p}$   & 
{$r_{\rm p}$ 
\footnote{Pericenter distance in units of $R_{\rm d}$ (=17.5 kpc)}} &
${\theta}_{1}$  &  
${\theta}_{2}$  &  
${\phi}_{1}$  &  
${\phi}_{2}$  &  
{$a_{\rm mrc}/a_{\rm mpc}$%
\footnote{Scale length ratio of MRCs to MPCs.}} &
{$t_{\rm v}$%
\footnote{The ratio of initial kinematic (either rotational or random)
energy to potential energy in a multiple merger.}} &
{$R_{\rm e}$%
\footnote{The half-mass radius of stars in simulation units of $R_{\rm d}$ (=17.5 kpc).
The real scale of the effective radius  of  an elliptical with $L_{\rm B}$ is 
$17.5 {(\frac{L_{\rm B}}{1.2 \times 10^{10} {\rm L}_{\rm B,\odot}})}^{0.5}
\times R_{\rm e}$ (listed in the table)
kpc.}}  &
{$R_{\rm e,mpc}$%
\footnote{The half-mass radius of MPCs in simulation units of $R_{\rm d}$ (=17.5 kpc).
The real scale of this can be estimated in the same way as the above for
$R_{\rm e}$.}} &
{$R_{\rm e,mrc}$%
\footnote{The half-mass radius of MRCs in simulation units of $R_{\rm d}$ (=17.5 kpc).
The real scale of this can be estimated in the same way as the above for 
$R_{\rm e}$.}} &
Comments \\
PM1 & 1.0 & 1.0 & 1.0 & 30 & 60 & 90 & 0 & 0.5 & -- & 0.37 & 0.96 & 0.53 & fiducial  \\
PM2 & 1.0 & 1.0 & 1.0 & 0 & 30 & 0 & 0 & 0.5 & -- & 0.39 & 0.98 & 0.50 & prograde-prograde  \\
PM3 & 1.0 & 1.0 & 1.0 & 150 & 180 & 0 & 0 & 0.5 & -- & 0.36 & 0.93 & 0.50 & retrograde-retrograde  \\
PM4 & 1.0 & 1.0 & 0.2 & 30 & 60 & 90 & 0 & 0.5 & -- & 0.39 & 1.00 & 0.53 & smaller angular momentum  \\
PM5 & 1.0 & 0.7 & 1.0 & 30 & 60 & 90 & 0 & 0.5 & -- & 0.33 & 0.85 & 0.47 & elliptic orbit  \\
PM6 & 1.0 & 1.0 & 1.0 & 0 & 150 & 90 & 0 & 0.5 & -- & 0.37 & 0.93 & 0.51 &  prograde-retrograde \\
PM7 & 0.1 & 1.0 & 0.5 & 0 & 30 & 0 & 0 & 0.5 & -- &  0.31 & 0.62 & 0.34 &  \\
PM8 & 0.3 & 1.0 & 0.5 & 0 & 30 & 0 & 0 & 0.5 & -- &  0.36 & 0.81 & 0.42 & \\
PM9 & 1.0 & 1.0 & 0.5 & 0 & 30 & 0 & 0 & 0.5 & -- &  0.49 & 1.15 & 0.59 & \\
PM10 & 0.1 & 1.0 & 0.5 & 0 & 30 & 0 & 0 & 0.1 & -- & 0.32 & 0.63& 0.10 &  \\
PM11 & 0.3 & 1.0 & 0.5 & 0 & 30 & 0 & 0 & 0.1 & -- & 0.37 & 0.90 & 0.17 &  \\
PM12 & 1.0 & 1.0 & 0.5 & 0 & 30 & 0 & 0 & 0.1 & -- & 0.50 & 1.09 & 0.13  \\
MM1 & 1.0 & --  & --  &  --  & -- & --  & -- &  0.5  &  0.50 & 0.77 & 1.31 & 0.95 &
{dispersion supported
\footnote{The kinematic energy of a multiple merger  is due totally 
to random motion of the five constituent galaxies.}} \\ 
MM2 & 1.0 & --  & --  &  --  & -- & --  & -- &  0.5  &  0.75 & 0.75 & 1.45 & 0.98 & 
dispersion supported \\
MM3 & 1.0 & --  & --  &  --  & -- & --  & -- &  0.5  &  0.25 & 0.73 & 1.21 & 0.87 &
dispersion supported \\
MM4 & 1.0 & --  & --  &  --  & -- & --  & -- &  0.5  &  0.50 & 0.64 & 1.08 & 0.75 &
{rotation supported
\footnote{The kinematic energy of a multiple merger  is due totally 
to rotational motion of the five constituent galaxies.}} \\ 
MM5 & 1.0 & --  & --  &  --  & -- & --  & -- &  0.5  &  0.75 & 0.69 & 1.17 & 0.79 & 
rotation supported \\
MM6 & 1.0 & --  & --  &  --  & -- & --  & -- &  0.5  &  0.25 & 0.61 & 1.01 & 0.73 &
rotation supported \\
\end{tabular}
\end{minipage}
\end{table*}

\section{The merger model}

\subsection{Progenitor spiral galaxies}

Since the numerical methods and techniques we employ for modeling
dynamical evolution of galaxy mergers have already been detailed 
elsewhere (Bekki \& Shioya 1998), we give only  a brief review here. 
The progenitor disk galaxies that take part in a merger are given 
a dark halo, a bulge, a stellar halo,  a thin exponential disk,
MPCs, and MRCs.
The total disk mass and size are $M_{\rm d}$ and $R_{\rm d}$, respectively. 
Henceforth, all masses are measured in units of
$M_{\rm d}$ and  distances in units of $R_{\rm d}$, unless otherwise specified. 
Velocity and time are measured in units of $v$ = $ (GM_{\rm d}/R_{\rm d})^{1/2}$ and
$t_{\rm dyn}$ = $(R_{\rm d}^{3}/GM_{\rm d})^{1/2}$, respectively,
where $G$ is the gravitational constant and assumed to be 1.0
in the present study. 
If we adopt $M_{\rm d}$ = 6.0 $\times$ $10^{10}$ $ \rm M_{\odot}$ and
$R_{\rm d}$ = 17.5\,kpc as fiducial values, then $v$ = 1.21 $\times$
$10^{2}$\,km\,s$^{-1}$  and  $t_{\rm dyn}$ = 1.41 $\times$ $10^{8}$ yr.

We adopt the density distribution of the NFW 
halo (Navarro, Frenk \& White 1996) suggested from CDM simulations:

 \begin{equation}
 {\rho}(r)=\frac{\rho_{0}}{(r/r_{\rm s})(1+r/r_{\rm s})^2},
 \end{equation} 

 where  $r$, $\rho_{0}$, and $r_{\rm s}$ are
the spherical radius,  the central density of a dark halo,  and the scale
length of the halo, respectively.  
The dark matter to disk mass ratio is fixed at 9 for all models.  
The value of $r_{\rm s}$ (typically $\sim$ 3$R_{\rm d}$) is chosen such that
the rotation curve of the disk is reasonably consistent with
observations for a given bulge mass. The $R^{1/4}$  bulge has a
mass of 0.17 and a scale length of 0.04,
both of which are consistent with observations for the Galactic
bulge (e.g., van den Bergh 2000).

The radial ($R$) and vertical ($Z$) density profiles 
of the  disk are  assumed to be
proportional to $\exp (-R/R_{0}) $ with scale length $R_{0}$ = 0.2 
and to  ${\rm sech}^2 (Z/Z_{0})$ with scale length $Z_{0}$ = 0.04
in our units, respectively.
In addition to the rotational velocity attributable to the gravitational
field of the disk and halo components, the initial radial and azimuthal velocity
dispersions are added to the disk component in accordance with
the epicyclic theory, and with a Toomre parameter value of $Q$ = 1.5
(Binney \& Tremaine 1987).
The vertical velocity dispersion at a given radius 
is set to be half as large as the radial velocity dispersion at that point, 
as is consistent with the trend observed in the Milky Way (e.g., Wielen 1977).

The radial scale length of a disk and 
the maximum rotational velocity for the adopted mass profiles
of dark matter, bulge, and disk is 3.5 kpc and 220 km s$^{-1}$,
respectively,  for the spiral  
with  $M_{\rm d}$ = 6.0 $\times$ $10^{10}$ $ \rm M_{\odot}$ and
$R_{\rm d}$ = 17.5\,kpc. The total mass of a disk galaxy is $10.17 \times M_{\rm d}$
corresponding to  $6.1 \times$ $10^{11}$ $ \rm M_{\odot}$
for all models. These adopted values are consistent with
those observed for the Galaxy (e.g., Binney  \& Tremaine 1987).
Thus total mass of a merger remnant is $10.17 (1+m_{2}) M_{\rm d}$  
(where $m_{2}$ is the mass ratio of merging two spirals) for a pair
merger and $50.85  M_{\rm d}$ for a multiple merger. The details
of these two different sets of merger models are described later.

\subsection{Initial distributions and kinematics of MPCs and MRCs}

The Galactic GCS and the stellar halo
have similar radial density profiles of ${\rho}(r)$ $\propto$ $r^{-3.5}$
(van den Bergh 2000).
We therefore assume that a GCS in a spiral has the following
radial density profile:

 \begin{equation}
 {\rho}(r)=\frac{\rho_{gc,0}}{({a_{\rm gc}}^2+r^2)^{1.75}},
 \end{equation} 

where  $r$, $\rho_{\rm gc, 0}$, and $a_{\rm gc}$ are
the spherical radius,  the central number density of the GCS,  and the scale
length of the GCS, respectively.  
We adopt different values of $a_{\rm gc}$ for MPCs and MRCs
where $\rho_{\rm gc, 0}$ is determined according to the adopted $a_{\rm gc}$.
The scale length is represented by $a_{\rm mpc}$ for MPCs
and $a_{\rm mrc}$ for MRCs and the the ratio of  $a_{\rm mrc}$
to $a_{\rm mpc}$ is set to be 0.5 for most of models.
This adopted value of 0.5 is reasonably consistent with
observations (Forbes et al. 1997).
The GCS with $a_{\rm mrc}$ $\sim$ $0.86R_{\rm d}$  
is consistent with the observed GC distribution of the Galaxy.
The stellar halo in  a spiral has the same distribution  as MPCs
in all models. 
The cut off radius ($R_{\rm c}$) beyond which no GC (and stellar halo)  particles are initially allocated
is set to be $2R_{\rm d}$ for MPCs and $R_{\rm d}$ for MRCs in
the models with $a_{\rm mrc}/a_{\rm mpc}$ = 0.5.

The GCS in a spiral is assumed to be supported purely by velocity dispersion
and its dispersion is assumed to be isotropic.
We therefore estimate the velocities of  GC particles 
from the gravitational potential
at the positions where they are located.

In detail, we first calculate the one-dimensional isotropic dispersion according to
the (local) virial theorem:

\begin{equation}
{\sigma}^{2}(r)=-\frac{U(r)}{3},
\end{equation}

where $U(r)$ is the gravitational potential at the position $r$.
Then we allocate a velocity to  each GC particle 
so that the distribution of velocities
of these particles has a Gaussian form with a dispersion equal
to ${\sigma}^{2}(r)$.

Note that we use the above $U(r)-{\sigma}^{2}(r)$ relation
rather than the  Jeans equation 
for a spherically symmetric system (Binney \& Tremaine 1987);

\begin{equation}
\frac{d(\rho (r){\sigma}^{2}(r))}{dr}=-\rho (r) \frac{d \Phi (r)}{dr}.
\end{equation}

This is firstly because the self-gravitating systems modeled in the present
study are composed of six different stellar components (dark matter halo,
$R^{1/4}$-law bulge, stellar halo, disk, MPC, and MRC) with non-analytical
radial density distributions and secondly because we introduce a cut off 
radius for each component.  
The derived ${\sigma}^{2}(r)$ at each location of a particle is 
consistent with those derived in the Jeans equation.

The total number of GCs ($N_{\rm gc}$) for each spiral is 1000
(MPCs + MRCs). This is several times larger than those observed
for the Galaxy or M31, and we adopt this number to improve
the statistics on the measured kinematics.
For example, the uncertainty in rotational velocity ($V_{\rm rot}$)
in each radial bin is estimated as $V_{\rm rot}/\sqrt{2(N_{i}-1)}$,
where $N_{i}$ is the total number of particles in each  bin.
This error is typically less than $\sim$ 20 km s$^{-1}$ in radial bins
for the outer part of a merger remnant
($R$ $\sim$ 20 kpc)  in the models with  $N_{\rm gc}$ = 1000.
This uncertainty is smaller than $V_{\rm rot}$  
($\sim$ 60 km s$^{-1}$) such that we can place reasonable constraints
on the rotational  kinematics of GCs in merger remnants.  
For models with $N_{\rm gc}$ = 100, the error bar is comparable
to $V_{\rm rot}$ so that it is at best challenging to derive any meaningful
conclusions on GC kinematics. 
Clearly, we could have derived radial profiles of $V_{\rm rot}$
with better statistics for models with, for example, $N_{\rm gc}$ = 10000.
However, such a GCS is consistent with only the 
most luminous cD galaxies. Thus 1000 GCs is a reasonable 
compromise between these two extremes.

Also, we stress here that
we assume that both MPCs and MRCs are dynamically supported
by velocity dispersion only: We do not intend to investigate
the influence of initial rotation of GCSs on the final kinematics
of GCSs in merger remnants.  
Statistical studies of the rotational properties of GCSs, in particular,
MRCs, have not yet  been performed for spiral galaxies.
Given the fact that some  MRCs in the Galaxy and M31 show rotational
kinematics (e.g., Huchra et al. 1991), this assumption
could be questionable for MRCs.
Suffice to say that we leave investigations of
the importance of initial rotation in GCSs to future studies.

 
The present study considers only the physical role of galaxy merging
in determining kinematical properties of GCSs in an idealized
manner, and does not attempt to address the origin of
GC kinematics and $S_{\rm N}$ of  giant ellipticals (e.g., M87 and NGC 1399)
in the central regions of galaxy clusters.
In such environments, accretion of GCs from cluster member galaxies
can play a significant role in determining physical properties of GCSs
(e.g., Bekki et al. 2003), and initial conditions based upon
cosmological simulations are probably more appropriate.
Cosmological simulations of GC formation in the early universe are still
at the level where individual GCs are unresolved (e.g., Kravtsov \& Gnedin  2003) 
such that there are still significant uncertainties
in the predicted properties of GCSs of the progenitors
of elliptical galaxies.
Thus we consider our approach to be complementary
to the current cosmological simulations.


\begin{figure}
\psfig{file=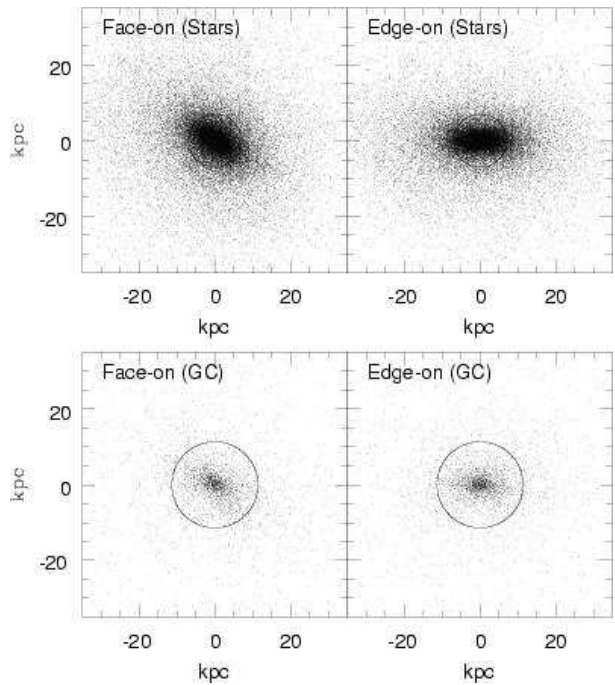,width=8.cm}
\caption{
Stellar (upper panels) and globular cluster (lower panels) distributions
projected onto the $x$-$y$ plane (i.e., face-on view, left)
and the $x$-$z$ plane (i.e., edge-on view, right)
for the merger remnant of the fiducial (PM1) model at $T$ = 4.5 Gyr,
where $T$ is time that has elapsed since the simulation starts.
GCs in the lower two panels include both MPCs and MRCs and
the circle in each panel represents the effective radius of
each GC component.  
}
\label{Figure. 1}
\end{figure}

\begin{figure}
\psfig{file=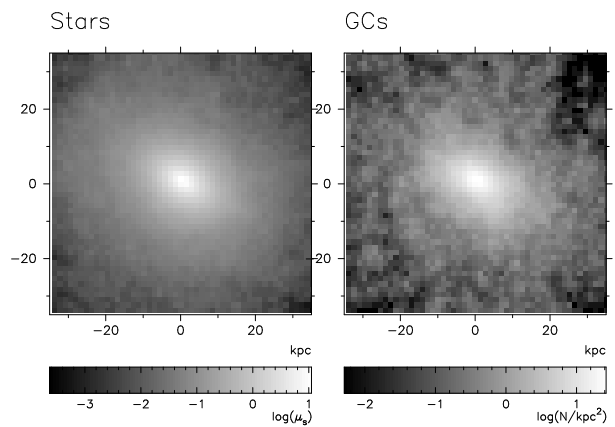,width=8.cm}
\caption{
The 2D smoothed density distributions of stars (left) and 
GCs (right) projected onto the $x$-$y$ plane
for the fiducial model at $T$ = 4.5 Gyr. 
Both stellar and GC distributions appear to be similarly flattened. 
}
\label{Figure. 2}
\end{figure}

\subsection{Orbital configurations}

\subsubsection{Pair mergers}

We investigate both a ``pair merger'' involving only two spiral galaxies
and a ``multiple merger'' containing five galaxies.
The multiple merger model for elliptical galaxy formation has been 
proposed (e.g.,  Barnes 1989; Weil \& Hernquist 1994, 1996;
Bekki 2001) 
to provide an evolutionary link between
a small compact group of galaxies and an isolated giant elliptical. 
For the pair merger model, the mass ratio of the two spirals ($m_2$) is
assumed to be a free parameter.
In all of the simulations of pair mergers, the orbit of the two disks is set to be
initially in the $xy$ plane and the distance between
the center of mass of the two disks
is  assumed  to be 6 in our units (corresponding to 105\,kpc). 
The pericenter distance ($r_{\rm p}$) and the eccentricity ($e_{\rm p}$)
in a pair merger are assumed  to be free parameters that control
orbital energy and angular momentum of the merger. 
The spin of each galaxy in a merger
is specified by two angles $\theta_{i}$ and
$\phi_{i}$, where suffix  $i$ is used to identify each galaxy.
$\theta_{i}$ is the angle between the $z$ axis and the vector of
the angular momentum of a disk.
$\phi_{i}$ is the azimuthal angle measured from the $x$ axis to
the projection of the angular momentum vector of a disk onto the $xy$ plane.

\subsubsection{Multiple mergers}

Each multiple merger model  contains equal-mass  spiral galaxies
with random orientations of intrinsic spin vectors which are 
uniformly distributed within  a sphere of size $6R_{\rm d}$.
The most important parameter in this multiple merger model
is the ratio of the initial kinematic energy ($T_{\rm kin}$) of the merger to 
that of initial potential ($W$).
By varying this ratio  ($t_{\rm v}$; defined as $|2T_{\rm kin}/W|$) from 0.25 to 0.75,
we investigate how  $t_{\rm v}$ controls the final GC kinematics
of the merger remnants.
We show the results of two extreme cases: (1) where the initial kinetic energy
of a multiple merger is due entirely to the random motion of  
the five constituent galaxies (referred to as ``dispersion supported'')
and (2) where it is due entirely to
(rigid) rotational motion (``rotation supported'').
These two cases enables us to understand how
initial rotation (or dispersion) controls the resulting kinematical
properties of GCs in merger remnants.

In total we have investigated 55 models of pair mergers,
but only present the results of 18 models which 
illustrate representative results on GC kinematics.
The time taken for the progenitor spirals to completely merge and reach 
dynamical equilibrium is less than 16.0 in our units ($\sim$ 2.2\,Gyr) for most of
our major merger models. 
The total number of particles is 112680 for a pair merger
and 281700 for a multiple merger. 
Table 1 summarises the model parameters for each  model:

Model number (column 1), the mass ratio $m_{2}$ (2),
the orbital eccentricity $e_{\rm p}$ (3),
the pericentre distance $r_{\rm p}$ (4),
$\theta_{1}$ (5), $\theta_{2}$ (6), $\phi_{1}$ (7), $\phi_{2}$ (8),  
$a_{\rm mrc}/a_{\rm mpc}$ (9), $t_{\rm v}$ (10), 
$R_{\rm e}$ (11), $R_{\rm e,mpc}$ (12), $R_{\rm e,mrc}$ (13),  and
comments (14). 
The method of  how to derive the half-mass radii of stars
($R_{\rm e}$), MPCs ($R_{\rm e,mpc}$), and MRCs ($R_{\rm e,mrc}$)
of the merger remnants
are described in \S 2.5.

Although we adopt a larger number (2000) of GCs in merger progenitor spirals
for the purpose of kinematical analysis with smaller error bars,
the present results of kinematical properties do not depend
on the number of initial GCs for $N_{\rm GC} \ge 200$ (Note here
that the error bars become significantly large 
for small $N_{\rm GC} \sim 200$). As shown later, GCSs both in pair mergers 
and in multiple ones can have a significant amount of rotation in
their outer parts. This is not due to the adopted large number of 
GCs and thus suggests that most merger remnants can have GCSs with
a significant amount of rotation.

\subsection{Methods for kinematical analysis}
\subsubsection{Radial gradients of $V_{\rm rot}$ and $\sigma$}

In estimating radial profiles of $V_{\rm rot}$ and $\sigma$ for a merger
remnant projected onto a given plane (e.g., the $x$-$y$ plane),
we first rotate the remnant such that
the major axis of the rotated stellar remnant coincides with the $x$-axis
(or $y$-axis) in the projection. 
Then we estimate $V_{\rm rot}$ and $\sigma$ of GCs at each radius along
the major axis in each projection to obtain the
radial profiles both for MPCs and MRCs
In estimating $V_{\rm rot}$ and $\sigma$ at each radius, we
adopt the same method as used by Peng et al. (2004a)
with a slit width  of $0.34 R_{\rm d}$. 
Radial profiles of  $V_{\rm rot}$ and $\sigma$ are estimated
for $R$ $\le$ $6R_{\rm e}$ beyond which 
uncertainties become excessive due to finite sampling.

\subsubsection{Smoothed 2D density and velocity fields}

We investigate the projected two-dimensional (2D) density and velocity
distributions of a GCS in each merger model for $0$ $\le$ $R$ $\le$ $2R_{\rm e}$,
where $R_{\rm e}$ is the stellar effective radius of the merger remnant.
Since each GCS is composed of 1000 GCs, 
we need to use smoothing methods  
to draw smoothly changing 2D density and velocity fields
from these discrete data.
Similar to Peng et al. (2004a,b)  we use a 3D Gaussian kernel function 
with a smoothing length of 0.17 in our units (corresponding to 3 kpc)
to smooth the velocity (density) field around each GC particle.
We choose this smoothing length of 0.17, so that we can examine 
{\it global} 2D fields without losing resolution within $R_{\rm e}$.
The simulation field is divided into a 50 $\times$ 50  grid
and the line-of-sight-velocity and velocity dispersion are estimated
in each grid square.

\subsubsection{Spin axis misalignment}

We also examine possible kinematical differences between the dark matter halo,
stars, MPCs, and MRCs for each merger remnant elliptical.
In order to estimate kinematical differences, we first derive
the normalized spin (angular momentum) vector for each component: 

\begin{equation}
{\bf L_{\rm DM} } = \frac{1}{C_{\rm DM}} 
\sum_{i=1}^{N_{\rm DM }} {\bf  x_{\rm i}} \times {\bf  v_{i}} , 
\end{equation}

where $N_{\rm DM}$, ${\bf  x_{\rm i}}$, ${\bf  v_{i}}$, and
$C_{\rm DM}$ are the total number of dark matter particles,
the position vector of each dark matter particle, the velocity vector of the particle,
and the constant to normalize the spin vector (thus equivalent to 
$\rm  L_{\rm DM}$). The same procedure is used for estimating
${\bf  L_{\rm S} }$ (for stars), ${\bf  L_{\rm MPC} }$, 
and ${\bf  L_{\rm MRC} }$.  
Based on these normalised spin vectors, the misalignment angle
(${\Theta}_{\rm DM-MPC}$) between the dark matter halo and the MPCs
may be estimated with the following formula:

\begin{equation}
{\Theta}_{\rm DM-MPC} = \arccos({\bf  L_{\rm DM}} \cdot {\bf  L_{\rm MPC}}) 
\end{equation}

The same formula is used for estimating ${\Theta}_{\rm S-MRC}$
and  ${\Theta}_{\rm DM-MRC}$.

\subsection{Half-mass radii of MPCs and MRCs}

We investigate half-mass (or half-number) radii of MPCs and MRCs and
compare them with the effective (or half-mass) radii of stars in ellipticals. 
We estimate  $R_{\rm e}$, $R_{\rm e,mpc}$, and $R_{\rm e,mrc}$
for stellar and GC particles that are within a radius of
$R_{\rm cut}$ of a merger remnant in each model.
We adopt $2R_{\rm d}$ (=35 kpc) as a reasonable value for $R_{\rm cut}$
and the above effective radii depend only weakly on $R_{\rm cut}$
for $R_{\rm cut}>R_{\rm d}$. For example,
$R_{\rm e}$ is 0.37 (6.5 kpc) for $R_{\rm cut}=2R_{\rm d}$
and 0.29 (5.1 kpc) for for $R_{\rm cut}=R_{\rm d}$.

In order to compare these effective radii with
observations for ellipticals with {\it different luminosities},
the values of  $R_{\rm e}$, $R_{\rm e,mpc}$, and $R_{\rm e,mrc}$ are
given in dimensionless units in the Table 1. 
For convenience, we also give 
$r_{\rm e}$, $r_{\rm e,mpc}$, and $r_{\rm e,mrc}$ 
given in real scales (kpc) below.

By assuming a $B$-band mass-to-light ratio of 5 
for the merger progenitor disks and adopting the observed relation
between $R_0$ (disk scale length)
and $L_{\rm d}$ ($L_{\rm d} \propto {R_0}^2$; Freeman 1970),
where $L_{\rm d}$ is the $B$-band total luminosity of a disk,
we can derive the effective radius of stars ($r_{\rm e}$)  for  an elliptical with
the $B$-band luminosity of $L_{\rm B}$ in units of $L_{\rm B \odot}$
from the value ($R_{\rm e}$) listed in the Table 1
as follows:

\begin{equation}
r_{\rm e} = 17.5 {(\frac{L_{\rm B}}{1.2 \times 10^{10} {\rm L}_{\rm B,\odot}})}^{0.5}
\times R_{\rm e} \,  {\rm kpc}.
\end{equation}

For example,  the elliptical in  the fiducial model with 
$L_{\rm B} = 1.2 \times 10^{10} {\rm L}_{\rm B,\odot}$ 
(or $M_{\rm d} = 6.0 \times 10^{10} {\rm M}_{\odot}$)
has $r_{\rm e}$ (real scale) of 6.5 kpc.
$r_{\rm e, mpc}$ and $r_{\rm e, mrc}$ can be derived 
from $R_{\rm e, mpc}$ and $R_{\rm e, mrc}$ for a given
$M_{\rm d}$ of a model  in the same
as $r_{\rm e}$.
 
Future observational studies of the structural parameters of GCSs 
will allow $r_{\rm e, mpc}$ and $r_{\rm e, mrc}$ to be derived 
separately for nearby ellipticals,
thereby probing the relative distributions of MPCs and MRCs with respect
to stars of their host galaxies (e.g., Forte et al. 2005). Therefore, theoretical predictions
on $r_{\rm e, mpc}/r_{\rm e}$ and $r_{\rm e, mrc}/r_{\rm e}$
are useful for interpreting these observational results.

\begin{figure}
\psfig{file=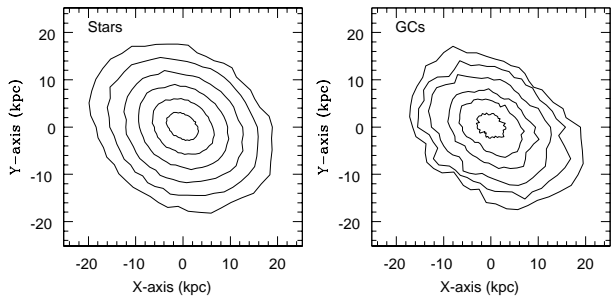,width=8.cm}
\caption{
Density contours for the smoothed density profile of stars (left) and 
GCs (right) shown in Fig. 2. 
Note that the major-axis of the stellar distribution is well aligned with that of
the GCS. The outer isophotes of the GCs become irregular
due to finite sampling.
}
\label{Figure. 3}
\end{figure}

\section{Results}

\subsection{The fiducial model}
\subsubsection{Spatial distributions of MPC and MRC}

Fig. 1 shows the final mass distributions of stars and
GCs of an elliptical galaxy
formed by major merging at $T$ = 4.5 Gyr in the fiducial
model. Both stars and GCs show flattened distributions 
in the face-on and the edge-on views, suggesting that 
the stars and GCs have intrinsically prolate-triaxial distributions.
This result demonstrates that galaxy major merging can transform
initially spherical distributions of GCs in spirals 
into moderately flattened distributions in remnant ellipticals.
The half-mass radius of the GCSs in each projected distribution is
a factor of 1.6 greater than that of the stars, and a factor of 2.6 smaller
than that of the dark matter halo. 
The half-mass radius of the GCs in the merger remnant elliptical  
increases by only $\sim$ 14\% with respect to that in the progenitor
spirals. This result implies that there may well be only marginal
differences between the half-mass radii of spirals and ellipticals
formed by major merging at a given luminosity.

\begin{figure}
\psfig{file=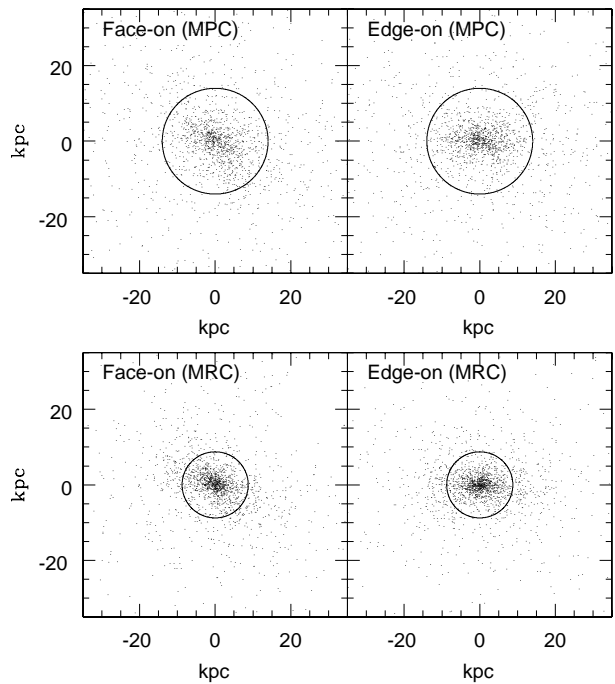,width=8.cm}
\caption{
The same as Fig. 1 but for MPCs (upper panels) and for MRCs (lower panels).
The major axis of the MPC distribution is well aligned with
that of the MRCs both for the face-on and the edge-on views.
This is an indication that intrinsic shapes of MPCs and MRCs are
quite similar to each other. Note also that MRCs show more compact distribution 
than MPCs.
}
\label{Figure. 4}
\end{figure}

Figs. 2 and 3 show, for the fiducial model, the smoothed distributions of stars and GCs
(projected onto the $x$-$y$ plane) and the isodensity contours 
of the distributions, respectively.
These two figures clearly indicate the alignment of the major axes
between the stellar distribution and the GCs.
The difference in position angles of the smoothed distributions
is only $\sim$ 10 degrees at the stellar effective radius ($R_{\rm e}$):
{\it The aligned major axis between stars and GCs appears to be 
one of the principal characteristics of  ellipticals
formed by major merging}.
The ellipticity of stars (${\epsilon}_{\rm s}$)
and GCs (${\epsilon}_{\rm gc}$)
at $R_{\rm e}$ is 0.18 and 0.22 respectively, 
indicating that the GC distribution is slightly more flattened
than the stellar one.
The stellar distribution shows a near constant ellipticity
($\sim$ $0.2$) for $R$ $>$ $R_{\rm e}$ whereas the GCs
show weak radial dependencies of the ellipticity, with
a slight increase ($\sim$ 0.3) at larger radii
($R$ = $3R_{\rm e}$). These results are due to the similarity 
in {\it intrinsic} mass distributions between the stars and GCs, 
such that they do not depend on viewing angle (i.e., projection). 
This alignment between stars and GCs is broadly consistent
with observations for nearby elliptical galaxies (e.g., Forbes et al. 1996).

Fig. 4 clearly demonstrates that 
(1) the 2D distributions of  MPCs and MRCs are similar to  each other
in the sense that both distributions are flattened along
the major axis of stars in both $x$-$y$ and $x$-$z$ projection,
(2) the major axes of the distributions of MPCs and MRCs are
nearly coincident with that of the stars,
and (3) MRCs have a more compact (by a factor of 1.6) distribution
than MPCs.
We confirm that major-axes alignment between stars, MPC, and MRCs
can be also seen.
The GC distribution is more flattened for MPCs 
(${\epsilon}_{\rm gc}$ $\sim$ 0.3)
than MRCs  (${\epsilon}_{\rm gc}$ $\sim$ 0.2)
within $R_{\rm e}$, although there is no significant difference
between the two outside $R_{\rm e}$.
Thus we conclude that the flattened spatial distributions 
of MPCs are expected to be characteristics of elliptical galaxies formed
in dissipationless major galaxy merging.

\begin{figure}
\psfig{file=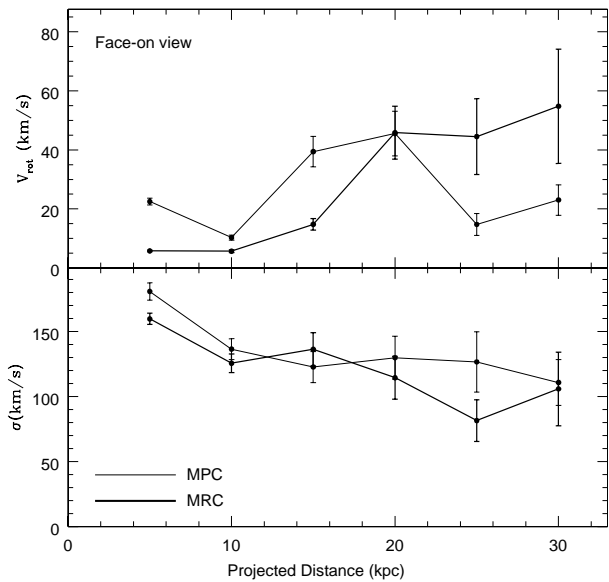,width=8.cm}
\caption{
Radial profiles of rotational velocity $V_{\rm rot}$ (upper panel)
and velocity dispersion $\sigma$ (lower panel) for MPCs (thin solid)
and for MRCs (thick solid) in the fiducial model projected onto
the $x$-$y$ plane (face-on view) at $T$ = 4.5 Gyr.
Both MPCs and MRCs show a significant rotation
for $R$ $>$ 20 kpc.
}
\label{Figure. 5}
\end{figure}

\begin{figure}
\psfig{file=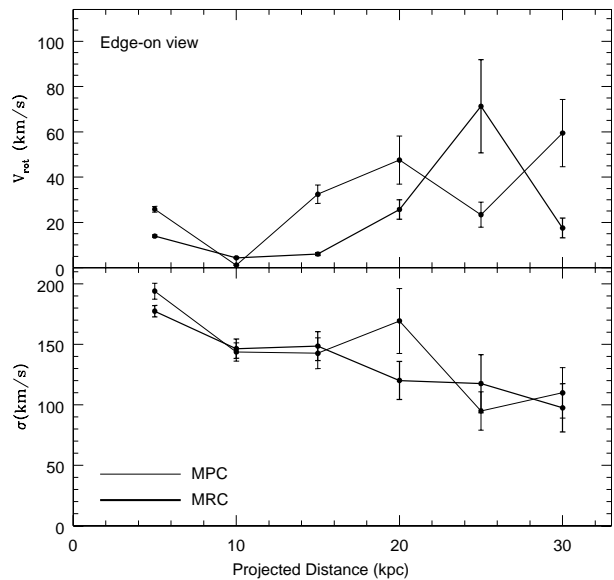,width=8.cm}
\caption{
The same as Fig. 5 but for the $x$-$z$ projection (edge-on view).
}
\label{Figure. 6}
\end{figure}

\subsubsection{Kinematics}

Fig. 5 shows the radial dependencies of rotational velocity
($V_{\rm rot}$) and velocity dispersion ($\sigma$) along
the major axis of the stellar distribution 
projected onto the $x$-$y$ plane (``face-on view'') 
for MPCs and MRCs in the fiducial model.
The rotation curves do not increase monotonically,
but $V_{\rm rot}$ at $R$ $\sim$ 20 kpc is significantly greater 
than that at $R$ $\sim$ 10 kpc for both GC subpopulations. 
Given the fact that GCs in the progenitor
spirals are given no net rotation initially,
these results demonstrate that outer GCs in the merger remnant 
ellipticals can exhibit rotation due to the redistribution of angular
momentum.
The amount of rotation in the GCs is relatively modest, with
$V_{\rm m}/{\sigma}_{0}$ = 0.25 for MPCs and 0.34 for MRCs.
Both MPCs and MRCs show radially decreasing profiles of $\sigma$
with no significant differences in the radial profiles between
the two GC populations.

As shown in Fig. 6, the simulated radial dependencies of $V_{\rm rot}$
are broadly similar between the two different
projections (i.e., the $x$-$y$ and the $x$-$z$ planes). 
The apparent independence of global $V_{\rm rot}$ profiles on
projection suggests that moderate rotation 
(30 $\sim$ 40 km s$^{-1}$) in the outer halo regions ($R$ $>$ 20 kpc)  
is a common characteristic of GC kinematics in 
major merger remnant ellipticals.

The radial profiles of $\sigma$ for MPCs do not significantly differ from that of 
MRCs, which probably reflects the fact that the two GC subpopulations
follow the same gravitational potential in the elliptical. 
$V_{\rm m}/{\sigma}_{0}$ and the slope of the $\sigma$  profile
for each GC population is slightly different between
the two different projections (e.g., $V_{\rm m}/{\sigma}_{0}$ = 0.25
for the $x$-$y$ projection and 0.3 for the $x$-$z$ one in MPCs).
The difference of the slopes of the $\sigma$  profiles are a result
of the anisotropic velocity dispersion profiles of the remnant.

Figs. 5 and 6 also show that the central velocity dispersion (${\sigma}_{0}$)
of MPCs is slightly higher (dynamically hotter) than that of
MRCs.  The  ratio of the dark matter to GCs ${\sigma}_{0}$ is
$\sim$ 1.05 (1.01) for MPCs and 1.20 (1.11)
for MRCs in the face-on (edge-on) view. 
This suggests that an estimation of the total mass
of the elliptical, by using the central velocity dispersion data of the GCs
and the virial (or Jeans) theorem, may lead to an underestimate of the total mass 
of the elliptical by up to 40 \%.
Thus, in principle, kinematical data of MPCs in an elliptical
allow for a more reliable mass estimation of the elliptical than MRCs.
The ${\sigma}_{0}$ of MPCs reflects that of the underlying dark matter
more effectively than  the MRCs.

Since the 2D velocity and velocity dispersion fields of the GCS
will be given in our forthcoming papers (Bekki et al. 2005), 
we here briefly summarize the main results of these.
Firstly, both minor axis and major axis rotation can be seen in
MPCs and MRCs. Secondly, the direction of the rotation along the major
axis for the dark matter halo is the same as those for MPCs and MRCs,
which suggests that observations of GC kinematics in an elliptical galaxy
can indirectly probe the kinematics of the dark matter halo.
Thirdly, the 2D $\sigma$ distribution appears to be more flattened
in MRCs than in MPCs, the direction of the flattening of MRCs
is the same as that of stellar density distribution
and appears to differ from that of MPCs,
and substructure is evident (i.e., local maxima of  $\sigma$)
for both MPCs and MRCs.

\subsubsection{Summary of generic results}

GCSs of merger remnants in pair merger models have the following 
generic properties: (1) well aligned major axes between stars, MPCs,
and MRCs, (2) a larger amount of intrinsic angular momentum in the
outer part of galaxies both for MPCs  and for MRCs,
(3) negative radial gradients of $\sigma$, (4) flattened 2D distributions
of velocity dispersion, and (5) rather complicated distributions of
line-of-sight-velocity fields.
Physical properties other than the above,  such as 
the ellipticity of 2D number  distributions, the slopes
of radial $V_{\rm rot}$ and
$\sigma$ profiles, and  the details of the velocity ellipsoids 
of GCSs, are model-dependent, as described below.

\begin{figure}
\psfig{file=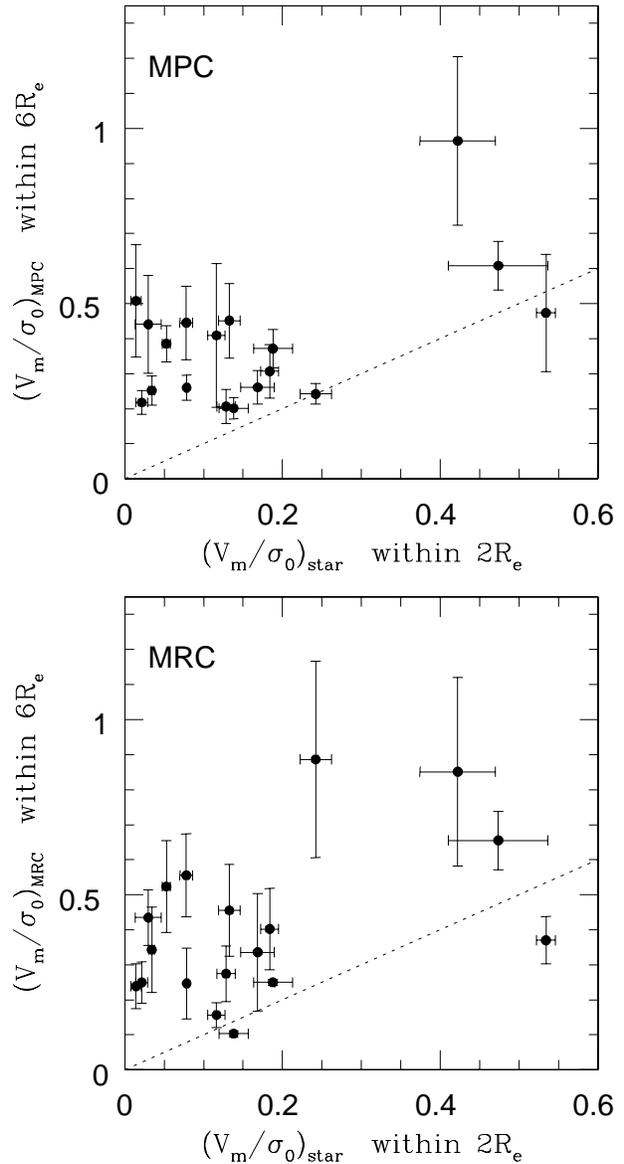,width=8.cm}
\caption{
Locations of six major merger models (PM $1-6$) on the
${(V_{\rm m}/{\sigma}_{0})}_{\rm star}$-${(V_{\rm m}/{\sigma}_{0})}_{\rm MPC}$
plane (upper panel) and on the
${(V_{\rm m}/{\sigma}_{0})}_{\rm star}$-${(V_{\rm m}/{\sigma}_{0})}_{\rm MRC}$
plane (lower panel).
Here three results for three different projections are derived for
each merger model so that 18 results are shown in each panel.
${(V_{\rm m}/{\sigma}_{0})}_{\rm star}$ is estimated for $R$ $\le$ $2R_{\rm e}$
whereas ${(V_{\rm m}/{\sigma}_{0})}_{\rm MPC}$
and ${(V_{\rm m}/{\sigma}_{0})}_{\rm MRC}$ are estimated for
$R$ $\le$ $6R_{\rm e}$.
The dotted line draws a boundary above which
${(V_{\rm m}/{\sigma}_{0})}_{\rm MPC}$ (or ${(V_{\rm m}/{\sigma}_{0})}_{\rm MRC}$) 
is larger than ${(V_{\rm m}/{\sigma}_{0})}_{\rm star}$.
}
\label{Figure. 7}
\end{figure}

\begin{figure}
\psfig{file=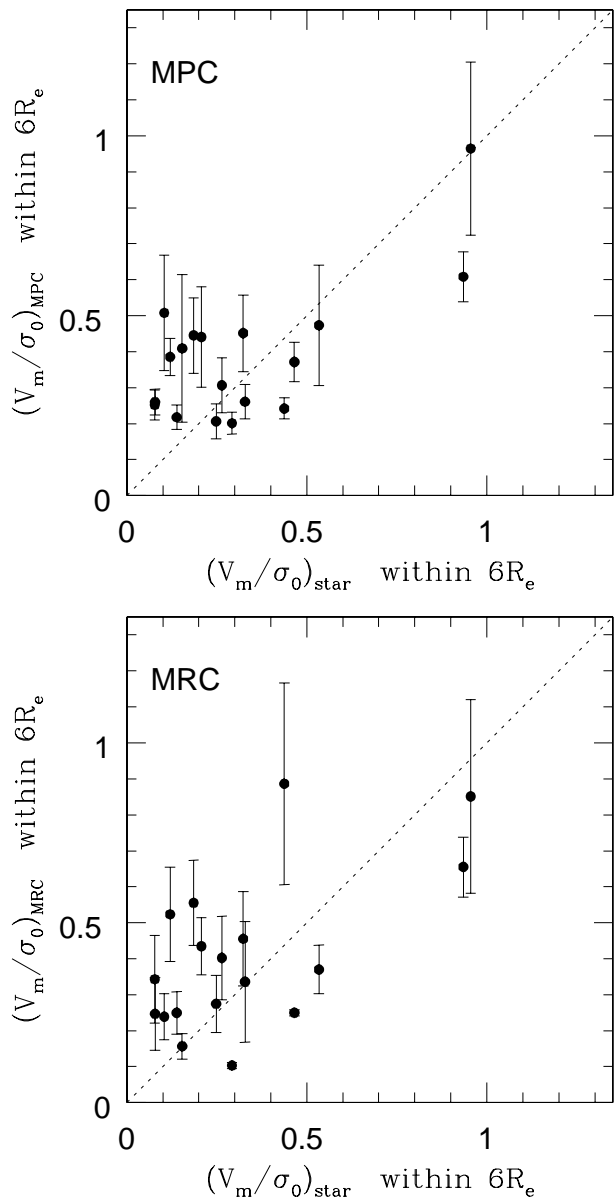,width=8.cm}
\caption{
The same as Fig. 7 but for ${(V_{\rm m}/{\sigma}_{0})}_{\rm star}$ 
estimated for $R$ $\le$ $6R_{\rm e}$. 
}
\label{Figure. 8}
\end{figure}

\begin{figure}
\psfig{file=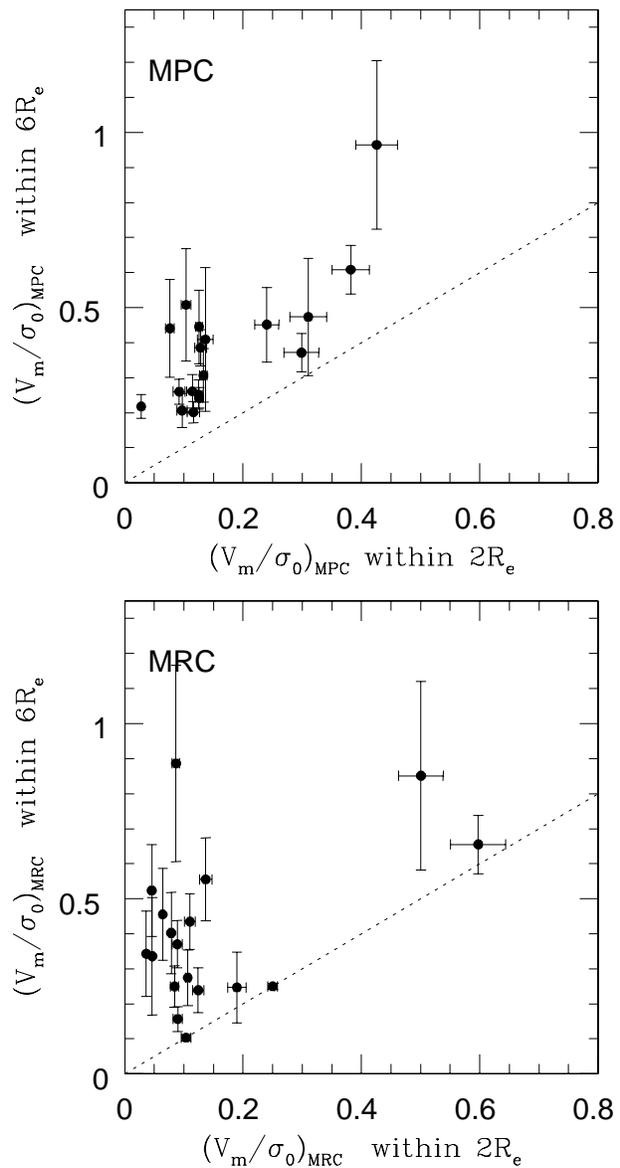,width=8.cm}
\caption{
Correlations between ${(V_{\rm m}/{\sigma}_{0})}_{\rm MPC}$ 
estimated for $R$ $\le$ $2R_{\rm e}$ and
those for $R$ $\le$ $6R_{\rm e}$ (upper)
and between ${(V_{\rm m}/{\sigma}_{0})}_{\rm MRC}$
estimated for $R$ $\le$ $2R_{\rm e}$ and
those for $R$ $\le$ $6R_{\rm e}$ (lower).
These correlations are derived from 18 results of 6 major merger
models with three different projections.
}
\label{Figure. 9}
\end{figure}

\subsection{Parameter dependencies}

\subsubsection{Orbital configurations of major merging}

It is important to ask: how do different orbital configurations 
effect the final kinematic properties of the GC system?
In the following, we summarize these dependencies. \\

(1) Irrespective of initial orbits,
both MPCs and MRCs in the merger remnant ellipticals show 
moderate amounts of rotation in their outer halo regions ($R$ $>$ 20 kpc).
$V_{\rm rot}$($r$), 
${\sigma}_{0}$, $V_{\rm m}$, and $V_{\rm m}/{\sigma}_{0}$
all depend on the orbital configuration.
Radial $V_{\rm rot}$ profiles can differ between MPCs and MRCs 
for a given projection. 

(2) All major merger models show $\sigma$ decreases
as a function of radius,  with the slope depending on projection and orbital
configuration. There are no significant differences in
the $\sigma$ profiles between MPCs and MRCs in our models.

(3) MPCs and MRCs generally show both major- and minor-axis
rotation in the 2D $V_{\rm los}$ distributions. MPCs and MRCs in most models
show quite flattened 2D $\sigma$ distributions, which reflect the fact  
that for most merger remnants, the GCS have anisotropic velocity dispersions.

(4) Both MPCs and MRCs in ellipticals formed from 
major merger models with prograde-prograde orbital configurations
show {\it figure rotation} with slow pattern speeds. 
Since dark matter halos in these models also show figure
rotation, this suggests that
kinematical studies of MPCs and MRCs in ellipticals could in 
principle test for global figure rotation in their dark matter halos.

(5) $V_{\rm m}/{\sigma}_{0}$ of MPCs differs
for models with different orbital configurations,
In addition, $V_{\rm m}/{\sigma}_{0}$ can be significantly  different
between stars, MPCs, and MRCs in a given galaxy. 
For example, Fig. 7 shows that $V_{\rm m}/{\sigma}_{0}$ of MPCs and MRCs in ellipticals 
are significantly larger than that of the main stellar bodies (within 2$R_{\rm e}$)
in nearly all models. 
This is a natural result of the outer GCs containing a larger amount
of intrinsic angular momentum.

(6) Fig. 8 shows that the outer stellar halos ($R$ = $6R_{\rm e}$)
of ellipticals formed from 
major merging have nearly the same values of $V_{\rm m}/{\sigma}_{0}$
as those of MPCs and MRCs. This result implies that if
ellipticals are formed by major merging, there should be
no significant differences in $V_{\rm m}/{\sigma}_{0}$
between stellar halos and other kinematic tracers such as
PNe or GCs. 

(7) Fig. 9 demonstrates  that there is a weak correlation between
$V_{\rm m}/{\sigma}_{0}$ within $2R_{\rm eff}$ and
$6R_{\rm eff}$  for MPCs in the sense that MPCs with
larger $V_{\rm m}/{\sigma}_{0}$ within $2R_{\rm eff}$
show larger $V_{\rm m}/{\sigma}_{0}$ within $6R_{\rm eff}$.
If the main stellar body of a merger remnant can obtain
larger angular momentum during galaxy merging 
(thus larger $V_{\rm m}/{\sigma}_{0}$ within $2R_{\rm eff}$),
then the outer halo of the remnant can probably also obtain
larger angular momentum 
(thus larger $V_{\rm m}/{\sigma}_{0}$ within $6R_{\rm eff}$).
The weak trend seen in MPCs is not so clear in MRCs.  
$V_{\rm m}/{\sigma}_{0}$ is larger in 
the outer regions for both MPCs and for MRCs in nearly all merger
models.

(8) The major-axes of MPCs and MRCs nearly coincides with that of stars
in each of the major merger models. 
This suggests that future observations
on {\it global distributions of GCs throughout the entire halo
regions of elliptical galaxies} enable us to test the merger
scenario of elliptical galaxy formation.
Kinematics of MPCs and MRCs in each model is strikingly similar
to that of the dark matter halo, which strongly suggests that
GCs can be regarded as tracers of kinematics of dark matter halos.

\begin{figure}
\psfig{file=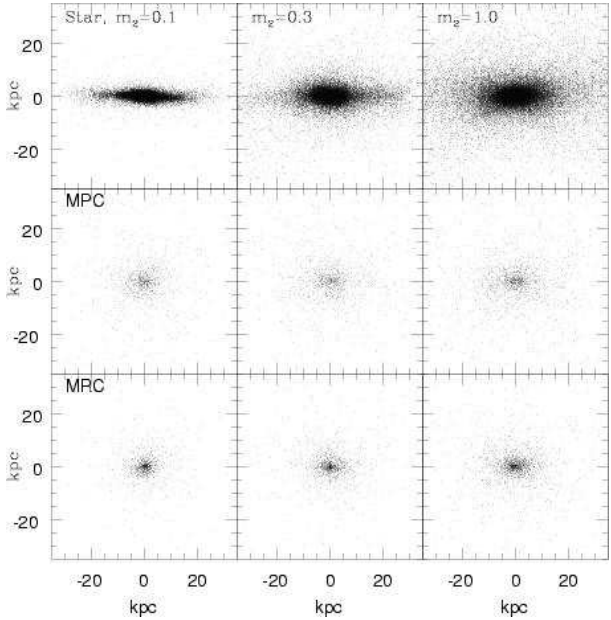,width=8.cm}
\caption{
Final spatial distributions of stars (top), MPCs (middle), and MRCs (bottom)
for three different merger models
(PM7, 8, and 9): $m_2$ = 0.1 (left), $m_2$ = 0.3 (middle),
and $m_2$ = 1.0 (right). 
}
\label{Figure. 10}
\end{figure}

\begin{figure}
\psfig{file=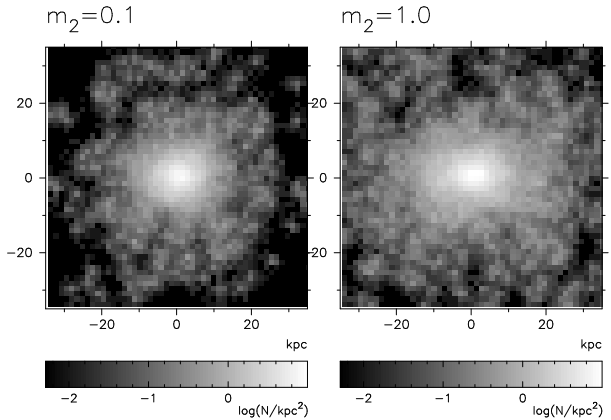,width=8.cm}
\caption{
Smoothed density distributions of MPCs projected onto the $x$-$z$ plane
for $m_2$ = 0.1 (left) and  $m_2$ = 1.0 (right).
}
\label{Figure. 11}
\end{figure}

\begin{figure}
\psfig{file=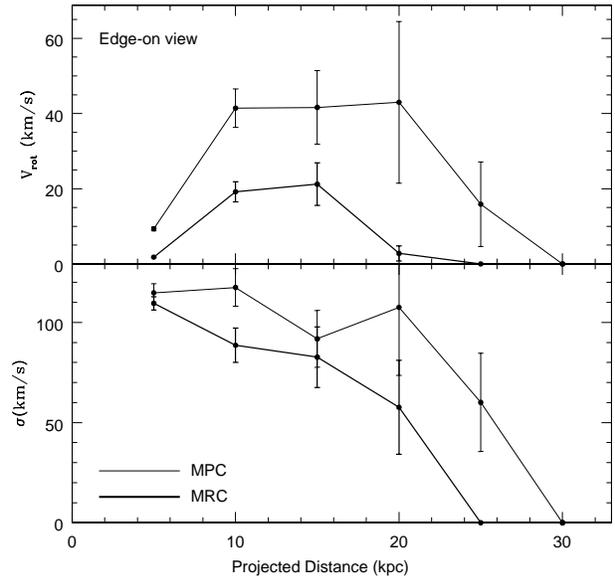,width=8.cm}
\caption{
The same as Fig.5 but for the model with $m_2$ = 0.1 (PM7).
}
\label{Figure. 12}
\end{figure}

\subsubsection{Mass ratios ($m_{2}$)}

The dependencies of kinematical properties
of MPCs and MRCs on the mass ratios ($m_{2}$)  of major merging
are summarized as follows. \\

(1) The spatial distributions of MPCs and MRCs are more flattened 
in models with larger $m_{2}$ for a given orbital configuration
and projection. For minor merger models with $m_{2}$ =0.1,
galaxy merging can not significantly transform an initially
spherical GC distribution into a flattened one due to
weak tidal perturbation. Figs. 10 and 11 show examples
of the dependence of final GC spatial distribution
on the mass ratio.
The ellipticity of the GCS  (${\epsilon}_{\rm gc}$) 
projected onto the $x$-$z$ plane at $R_{\rm e}$ is 0.08
for $m_{2}$ = 0.1 and 0.20 for $m_{2}$ = 1.0 for this nearly prograde-prograde 
configuration. 

(2) Stellar distributions of galaxies become more spherical
(i.e., smaller stellar ellipticity ${\epsilon}_{\rm s}$)
for larger $m_{2}$, whereas the GCSs become more flattened
(larger ${\epsilon}_{\rm gc}$). Therefore,
more flattened early-type galaxies (e.g., S0s) may show
smaller ${\epsilon}_{\rm gc}/{\epsilon}_{\rm s}$. 
It has been argued that minor or unequal-mass mergers of spirals 
can become S0s (e.g., Bekki 1998), and a statistical 
study of ${\epsilon}_{\rm gc}/{\epsilon}_{\rm s}$
can test this hypothesis.

(3) MRCs in flattened early-type galaxies formed by mergers with small $m_{2}$
($\sim$ 0.1) show a smaller amount of rotation 
(i.e., $V_{\rm m}/{\sigma}_{0}$  $\sim$ 0.1) when compared
with MPCs. Fig. 12 shows an example of this $V_{\rm rot}$ difference
between MPCs and MRCs in a  merger model with $m_{2}$ = 0.1 in which
the remnant looks like an S0. This difference is smaller for
models with larger $m_{2}$. These results imply that more flattened
early-type galaxies are likely to show larger $V_{\rm rot}$ 
in MPCs than in MRCs. 
Initial results for the GC kinematics of the S0 NGC 524 support the 
notion that these galaxy types may result from 
minor/unequal-mass merging of spirals (Beasley et al. 2004).

(4) The radial gradients of $\sigma$ are more likely to be steeper
for early-type galaxies formed by mergers with smaller $m_{2}$.
$V_{\rm rot}$  of MPCs and MRCs 
of E/S0 merger remnants with small $m_{2}$ 
reaches a maximum in the inner parts of their host galaxies. 
Independent of $m_{2}$, MPCs show larger velocity dispersions
than MRCs, which implies that MPCs in
early-type galaxies are {\it in general} dynamically hotter systems 
than MRCs.

\begin{figure}
\psfig{file=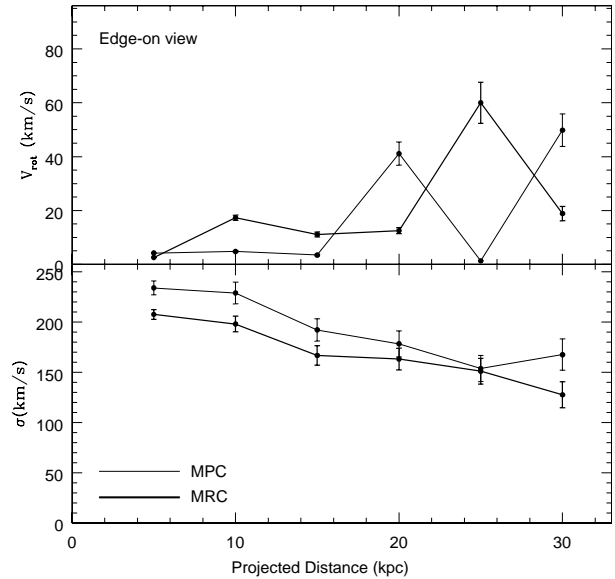,width=8.cm}
\caption{
The same as Fig.5 but for the multiple merger model MM1.
}
\label{Figure. 13}
\end{figure}

\begin{figure}
\psfig{file=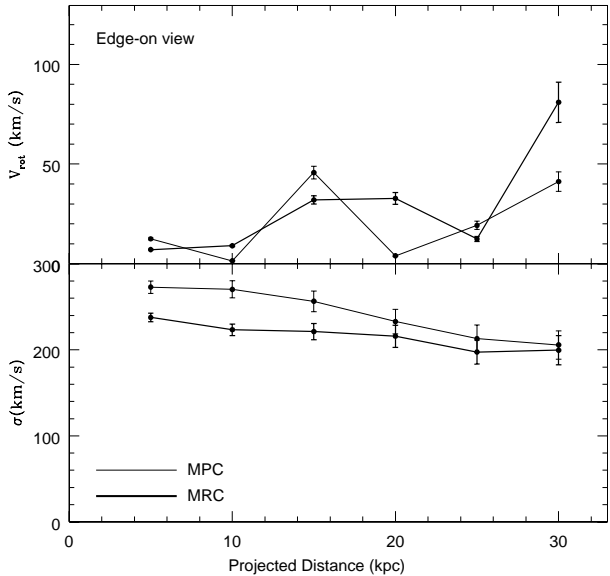,width=8.cm}
\caption{
The same as Fig.5 but for the multiple merger model MM4.
}
\label{Figure. 14}
\end{figure}

\subsubsection{Multiple mergers}

Most of the results for multiple merger models 
(MM $1-6$) are essentially
the same as those obtained for pair merger ones. We thus summarize
the salient dependencies in multiple merger models.

(1) Major-axis alignment between MPCs, MRCs, and stars in their
projected 2D distributions can be seen in all models.
The half-number radii of GCSs in ellipticals formed by multiple merging
are a factor of $\sim$1.3 larger than those formed by pair merging.
Both MPCs and MRCs show flattened 2D distributions in all multiple
merger models.

(2) Both MPCs and MRCs show significant amounts of rotation in 
the halo regions (20 $<$ $R$ $<$ 30 kpc) of their host galaxies.
Independent of model parameters, MPCs show larger $\sigma$ than
MRCs. Fig. 13 shows an example of the present model
showing these general trends in GC kinematics of ellipticals
formed by multiple merging. 
$V_{\rm m}/{\sigma}_{0}$ is generally larger in ellipticals
formed by multiple mergers initially having global rotation
(i.e., MM4, 5, and 6) than those initially having no net rotation 
(i.e., MM1, 2, and 3).

(3) Some remnant ellipticals show flattened radial ${\sigma}$ profiles
within $2R_{\rm e}$ ($\sim$ 20 kpc). Fig. 14
shows the result of the model MM4, in which both MPCs and MRCs
have quite flattened  ${\sigma}$ profiles.
These flattened  ${\sigma}$  profiles are not evident in pair merger models
and may be regarded as  one of the characteristics
of multiple merger models.
Such flattened ${\sigma}$ profiles seen 
in some Es (e.g., NGC 4472; C\^ote et al 2003) could be understood
in terms of multiple galaxy merging.

(4) Figs. 15 and 16 shows the radial $V_{\rm rot}$ and ${\sigma}$ profiles
of all pair and multiple merger models with three different projections,
respectively. Although the total mass of the merger remnants in pair
merger models are the same,  
{\it both the projected  $V_{\rm rot}$ and ${\sigma}$ profiles}
are quite diverse. This is true for the multiple merger models.
These results indicate that in deriving total masses of
ellipticals using GC kinematics (i.e.,  radial $V_{\rm rot}$ and ${\sigma}$ profiles)
and the Jeans equation (Binney \& Tremaine 1987),
more careful dynamical analysis and physical interpretations should be done.
In particular, the diverse $V_{\rm rot}$  profiles of GCSs 
imply that the contribution of global rotation should be included in
the mass estimation based on the Jeans equation.

(5) Figs. 15 and 16 also confirm that radial  ${\sigma}$ profiles
are in general flatter in multiple merger models than in pair mergers
both for MPCs and MRCs. Some pair and multiple merger models
show significant rotation in MPCs: $V_{\rm m}/{\sigma}_{0}$
ranges from 0.2 to 1.0.
In this regard, it is interesting to consider the Milky Way, 
in which the MPCs have a net rotation of $\sim$ 60 km s$^{-1}$
(Freeman 1985). This corresponds to $V_{\rm m}/\sigma_{0}$ = 0.47 
for $V_{\rm m}$ = 220 km s$^{-1}$ and $\sigma_{0}$ $\simeq$ $V_{\rm m}/\sqrt3$.
The above result suggests that some Es have MPCs that
rotate more rapidly than their counterparts in late-type spirals.

\begin{figure}
\psfig{file=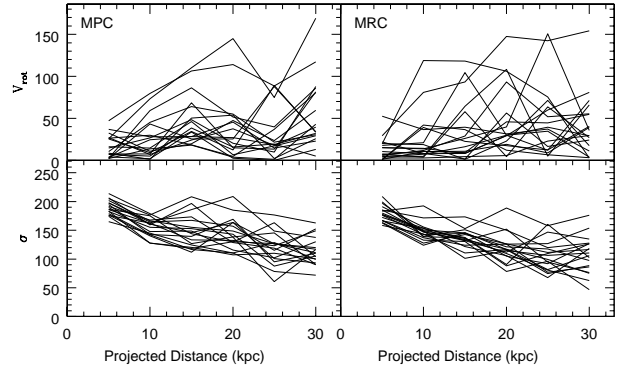,width=8.cm}
\caption{
Radial profiles of $V_{\rm rot}$ (upper two) and $\sigma$ (lower two)
for MPC (left) and MRC (right) in the six pair major merger models
(PM $1-6$) with three different projections. In total 18 model results
are given in each frame.
}
\label{Figure. 15}
\end{figure}

\begin{figure}
\psfig{file=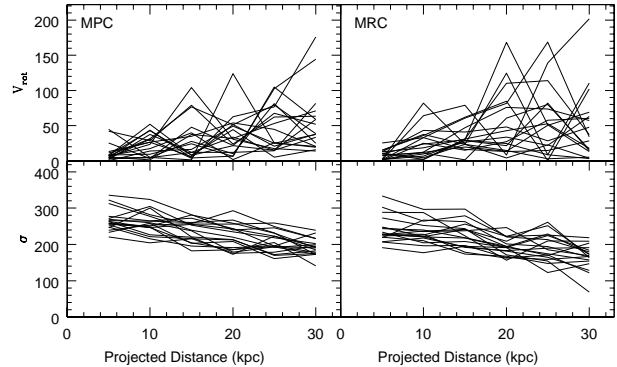,width=8.cm}
\caption{
The same as Figure 15 but for
the six multiple major merger models
(MM $1-6$) with three different projections. In total 18 model results
are given in each frame.
}
\label{Figure. 16}
\end{figure}

\begin{table*}
\centering
\begin{minipage}{185mm}
\caption{Characteristics of GCSs in early-type galaxies formed
from different types of galaxy mergers}
\begin{tabular}{|c||c||c||c|}  \hline 
Merger type & 
Host morphology   & 
GCS structure   & 
GCS kinematics  \\  \hline  
Major & E & More flattened  & Rotation at larger $r$ \\ \hline
Minor/unequal-mass & Flattened E/S0 & More spherical  & 
Little  rotation at larger $r$  \\ \hline
\end{tabular}  \\
\end{minipage}
\end{table*}

\section{Discussion}

\subsection{Kinematical differences of GCSs in cluster and field ellipticals}

Previous numerical studies have shown that tidal stripping of GCs
from cluster members, by the cluster global tidal
field and galaxy-galaxy interactions, are important
physical processes in the evolution  GCSs in cluster galaxies
(e.g., Muzzio 1987).
Recent more sophisticated numerical studies (Bekki et al. 2003)
have shown that more than 50 \% of GCs initially
within a cluster E can be stripped by cluster tidal fields
(resulting in a lower $S_{\rm N}$). Local $S_{\rm N}$ (i.e., $S_{\rm N}$
at a given radius) can be reduced even further due to efficient tidal stripping 
of GCs in the outer parts of Es.
The present study has demonstrated that GCs in the outer parts
of Es formed by major merging have a significant amount of
rotation and thus show large $V_{\rm m}/{\sigma}_{0}$. 

Therefore, the logical conclusion from the present and previous works 
is that GCSs of cluster Es are likely to have lower $V_{\rm m}/{\sigma}_{0}$
than those of field Es owing to the selective stripping of outer GCs in the cluster environment,
{\it if both Es were formed by major galaxy merging}.
Such differences are expected to be most clearly 
seen in field and cluster Es that are not located in 
the centers of their host clusters, since the GCSs
of cD galaxies may have been influenced by several ``secondary'' physical
processes such as GC accretion from cluster member galaxies and
minor merging of dwarfs with GCs, after their formation.

\subsection{Origin of S0s}

The question of how and when red S0 galaxies 
formed in the field and in clusters remains an
outstanding problem in astrophysics.
Morphological and spectroscopic observations
indicate that there is a smaller fraction of S0 galaxies 
in clusters of galaxies at $z\sim 0.4$
than locally (Dressler et al. 1997; Couch et al. 1998).
This, combined with the increased
fraction of  blue late-type spirals
in some  distant clusters (e.g., Butcher \& Oemler 1978),
suggest that some mechanism drives strong evolution  from blue spirals into
red S0s in the course of cluster evolution.

The proposed theoretical models for S0 formation are
ram pressure stripping (Farouki \& Shapiro 1980),
tidal compression by the gravitational field of clusters (Byrd \& Valtonen 1990),
tidal truncation  of gas replenishment (Larson, Tinsley, \& Caldwell 1980),
and galaxy mergers (Bekki 1998).
Although theoretical studies of the spectrophotometric evolution 
of S0s (e.g., Shioya et al. 2002, 2004) and on the morphological transformation
processes of spirals in clusters (Bekki et al. 2001, 2002)
have tried to identify the origin of cluster S0s,
it remains unclear which model can best explain the observations.

The present study suggests that GC kinematics of S0s can provide a fresh clue
to their origin. Our simulations have demonstrated that if S0s are formed
by minor/unequal-mass merging, then MPCs and MRCs will show different
kinematics when compared with those of their progenitor spiral.
In particular,  a significant  amount of rotation in MPCs 
and  larger $V_{\rm m}/{\sigma}_{0}$ in  MPCs than in MRCs
appears to be  principal characteristics of GCSs in S0s formed
by such merging. 
Other cluster-related physical processes such as ram pressure
stripping and tidal truncation of gas replenishment will not
change the outer dynamical properties of GCSs in S0 progenitor
spirals, so that GCSs of S0s formed entirely by these processes
will not show any net rotation.
It would also be difficult for the GCSs of the S0 progenitors to
be imparted with a net rotation if tidal compression by the 
gravitational field  of clusters transforms spirals into S0s.
Future kinematical studies of GCSs
in the field and in clusters can potentially unravel the postulated 
different formation processes between field and cluster S0s.

\section{Conclusions and Summary}

We have numerically investigated the kinematic properties of 
GCSs in E/S0s formed by dissipationless galaxy merging.
The dynamical evolution of metal-poor GCs (MPC) and metal-rich GCs (MPC), 
initially associated with the merger progenitor spirals, is followed
and their resulting kinematics examined. We summarize our principle result as follows.

(1) Both MPCs and MRCs are expected to have significant amounts 
of rotation in Es formed by major galaxy merging with 
mass ratios ($m_{2}$) $\sim$ 1. This remains true even if the MPCs and MRCs 
initially have no net rotation in the progenitor spirals.   
This arises because the GCs obtain angular momentum with respect
to their host galaxies during merging owing to the conversion 
of orbital angular momentum into intrinsic angular momentum of
the remnants, particularly in 
the outer regions of the mergers. 
Both MPCs and MRCs show positive radial gradients of
rotational velocity ($V_{\rm rot}$)
in the sense that outer parts of GCSs in Es ($R$ $>$ $2R_{\rm e}$)
show larger $V_{\rm rot}$ 
than the inner parts ($R$ $\sim$ $R_{\rm e}$). 

(2) MPCs show slightly larger central velocity dispersion than MRCs for
most major merger models, which indicates that MPCs are dynamically hotter stellar
systems than MRCs. Velocity dispersion is likely to be higher in MPCs than
in MRCs at any radius in ellipticals.

(3) $V_{\rm m}/{\sigma}_{0}$, where $V_{\rm m}$ and ${\sigma}_{0}$ 
are the maximum rotational velocity and the central velocity dispersion
of a GCS, respectively,
range from 0.2 to 1.0 for MPCs and from 0.1 to 0.9 for MRCs within
$6R_{\rm e}$ of Es formed by major merging. 
The distributions of $V_{\rm m}/{\sigma}_{0}$ are similar
between these two GC populations.
For most major merger models,
$V_{\rm m}/{\sigma}_{0}$ of GCSs within $6R_{\rm e}$ is greater
than that of stars of Es within $2R_{\rm e}$.
MPCs (and MRCs) with larger  $V_{\rm m}/{\sigma}_{0}$ 
for $R$ $\le$ $2R_{\rm e}$
show larger $V_{\rm m}/{\sigma}_{0}$  for $R$ $\le$ $2R_{\rm e}$.
This correlation in GC kinematics between the inner and outer 
regions of Es is a natural result of global dynamical relaxation
and angular momentum transfer, and in principle provides a 
test for the major merger scenario of elliptical galaxy formation.

(4) Radial profiles of $V_{\rm rot}$ 
for MPCs and MRCs are diverse,
and are sensitive to orbital configuration, $m_{2}$, and 
the viewing angle.
Radial profiles of $V_{\rm rot}$ can differ
between MPCs and MRCs which reflects the fact
that the dynamical evolution of GCs during major merging depend
on the GCs' initial distribution.
The negative radial gradients 
in velocity dispersion of MPCs and MRCs can also differ
for different model parameters.
However, there are no significant differences in the profiles
between MPCs and MRCs in the remnant elliptical.

(5) Outer MRCs in flattened early-type galaxies (e.g., S0s)
formed by merging with small mass ratios ($m_{2}$ $\sim$  $0.1$) 
do not show any significant rotation.
Furthermore,  $V_{\rm m}$ is likely to be larger in
MPCs than in MRCs for these flattened systems.
These results suggest that the kinematic properties of GCSs in 
S0s can be quite different from those in Es.
The $V_{\rm rot}$ in S0 galaxies obtain their  
peak values in their inner regions.
The fundamental characteristics of GCSs in E/S0s formed
from different mergers are summarized in  Table 2.

(6) 2D distributions of line-of-sight-velocities of GCSs can be useful
for examining global differences in kinematics between MPCs and
MRCs. Both MPCs and MRCs show rotation along the minor- and major-axes of 
their host E for most major merger models.
The general patterns of the 2D velocity fields are similar to
each other between MPCs and MRCs.

(7) 2D velocity dispersion fields of GCSs in Es formed
by major merging show flattening both in MPCs and in MRCs. This result is
a clear indication of an anisotropic velocity dispersion in the GCSs,
and demonstrates that major merging processes can change 
an initially isotropic velocity dispersion in the GCSs of spirals
into an anisotropic one in the GCSs of Es. 
The direction of flattening appears to be similar  
between the two GC populations in Es.

(8) Kinematic properties of GCSs in Es formed by multiple
major merging are generally similar to those of GCSs in
Es formed by pair merging. However, a significant difference in GC kinematics
between the two merger types is that 
very flattened radial $\sigma$ profiles
are only seen in  Es formed by multiple merging.
The major axis of the GCS distribution is nearly coincident with
that of the stellar distribution in an E formed from both pair 
and multiple mergers 

(9) Kinematic properties of MPCs in Es are similar to
those of the surrounding dark matter halos, which implies  
that MPCs are a good probe of 
the kinematics of dark halos.
This similarity in kinematics suggests that  dynamical processes
of major galaxy merging (i.e., violent relaxation and angular momentum
redistribution) can align the spin axis of the GCS with those
of the parent galaxy halo.  

\section{Acknowledgment}
We are  grateful to the anonymous referee for valuable comments,
which contribute to improve the present paper.
KB and DAF acknowledge the financial support of the Australian Research Council
throughout the course of this work.
The numerical simulations reported here were carried out on GRAPE
systems kindly made available by the Astronomical Data Analysis
Center (ADAC) at National Astronomical Observatory of Japan (NAOJ).
This work was supported by NSF grant AST 02-06139.\\

\end{document}